\documentclass[aps,prd,twocolumn,showpacs,amsmath,nofootinbib]{revtex4-1}

\usepackage{color}
\newcommand{\beqa}{\begin{eqnarray}}
\newcommand{\eeqa}{\end{eqnarray}}
\usepackage{float}
\usepackage{epsfig}
\usepackage{graphicx}
\usepackage{latexsym}
\usepackage{amsmath}
\usepackage{mathrsfs}
\usepackage{dcolumn}% Align table columns on decimal point
\usepackage{bm}%

\begin{document}

\title{Constraints on primordial black holes with CMB spectral distortions}

\author{Yupeng Yang}

\affiliation{School of Physics and Physical Engineering, Qufu Normal University, Qufu, Shandong, 273165, China}

\begin{abstract}
In the mixed dark matter scenarios consisting of primordial black holes (PBHs) and weakly interacting massive particles (WIMPs), WIMPs can be accreted onto PBHs to form ultracompact minihalos (UCMHs) with a density spike in the early universe. Compared with the classical dark matter halo, UCMHs are formed earlier and have a higher density of center. Since the annihilation rate is proportional to the squared number density of WIMPs, it is expected that WIMPs annihilation within UCMHs is enhanced and has influences on the early universe. Between the time of recombination and matter-radiation equality, the energy released from WIMPs annihilation within UCMHs is injected into the Universe resulting in CMB $y$-type distortion. We investigate these effects and derive the upper limits on the abundance of PBHs taking advantage of the observational results of Far Infrared Absolute Spectrophotometer (FIRAS). We find that for the WIMPs mass range $1\le m_{\chi}\le 1000~\rm GeV$, the upper limits on the abundance of PBHs are $5\times 10^{-3}\le \Omega_{\rm PBH}\le 5\times 10^{-2}$.

\end{abstract}

%\keywords{}
\maketitle

\section{introduction} 
Although the astronomical observations and related theoretical studies suggest that about $27\%$ of the Universe is dark matter (DM), the nature of them remains a mystery~\cite{Bertone:2004pz,Jungman:1995df,Arbey:2021gdg,planck-2018}. Among the DM candidates, weakly interacting massive particles (WIMPs) is the most competitive one~\cite{Bertone:2004pz,Jungman:1995df}. Although many experiments have been performed in order to find WIMPs, no confirmed signal has been found so far. Nevertheless, the properties of WIMPs can be constrained in light of the null detection results~\cite{Arcadi:2017kky,Fermi-LAT:2016uux,HESS:2016mib,MAGIC:2016xys}. Recently, the primordial black hole (PBH) dark matter model has received renewed interest especially due to the direct detection of gravitational waves~\cite{Bird:2016dcv,Deng:2021ezy,Franciolini:2021tla,Wang:2016ana,LIGO}. 

PBHs can be formed in the early universe via the collapse of large density perturbation~\cite{carr,pbhs_review}. The PBHs mass covers a wide range and their abundance can be constrained using different astronomical observations (see, e.g., Refs.~\cite{Carr:2020gox,Carr:2021bzv,Oncins:2022ydg} for a review, and Refs.~\cite{Boshkayev:2022vlv,Boshkayev:2022vlv,Kurmanov:2021uqv,Boshkayev:2021chc,Boshkayev:2020kle,DAgostino:2022ckg} for the accretion cases). By studying existing observational results, it has been found that PBHs in most mass ranges can only constitute part of DM~\cite{Carr:2021bzv}. For this case, we assume here that DM is mainly composed of two components. One is general particle dark matter, such as WIMPs, and the other is PBHs. In the following, we will use DM to represent WIMPs unless otherwise specified. Both theoretical research and numerical simulations have shown that due to the accretion of DM onto PBHs, ultracompact minihalos (UCMHs) can be formed around PBHs with a density profile $\rho_{\rm DM}(r)\sim \rho^{-9/4}$~\cite{0908.0735,Eroshenko:2016yve}. It is expected that the annihilation rate of DM particles within UCMHs is very large, because it is proportional to the square number density. The influences of DM annihilation within UCMHs on the astronomical observations have been investigated by previous works, while those studies have focused on the epoch after recombination $z\lesssim z_{\rm rec}\simeq 1100$~\cite{ucmh_scott,Lacki:2010zf,Josan:2010vn,dongzhang,Yang:2011jb,Bringmann:2011ut,Yang:2012qi,Yang:2017cjm,epjplus-2,Zhang:2021mth,Cheng:2020iym,Yang:2020zcu,Furugori:2020jqn,Kadota:2021jhg,Tashiro:2021xnj}. On the other hand, since UCMHs can be formed before cosmological recombination, DM annihilation within UCMHs should have influences on the early universe.

It has been pointed out that extra energy injection into the primordial plasma in the early universe can lead to the deviation of CMB from the  blackbody spectrum~\cite{McDonald:2000bk,Chluba:2011hw,waynehu,Chluba:2019nxa,Bernstein,Tashiro,Ricotti:2007au,Sunyaev:2013aoa,Chluba:2020oip,Kohri:2014lza,osti_6147408}. The CMB $y$-type and $\mu$-type distortion can be produced if extra energy is injected in the redshift range $1100\lesssim z\lesssim 5\times10^{4}$ and  $5\times 10^{4}\lesssim z \lesssim 2\times10^{6}$, respectively. The spectral distortion caused by the injection of extra energy released from PBHs or DM annihilation has been investigated in previous works~\cite{McDonald:2000bk,carr,Pani:2013hpa,Nakama:2017xvq,Tashiro:2008sf,Chluba:2011hw,PhysRevD.98.023001,Chluba:2019nxa}. In this paper, we will focus on the influences of DM annihilation within UCMHs on the CMB spectral distortion. Since the mass of a UCMH grows slowly before the epoch of matter-radiation equality ($z_{\rm eq}\simeq 3411$), we will focus on the redshift range $z_{\rm rec} \le z\le z_{\rm eq}$. That is, we will mainly investigate the influences of DM annihilation within UCMHs on the CMB $y$-type distortion. Based on the observations of the Far Infrared Absolute Spectrophotometer (FIRAS), current upper limits on the CMB $y$-type distortion is $|y|<1.5\times 10^{-5}$~\cite{Fixsen:1996nj}. By requiring that the CMB spectral distortion caused by DM annihilation within UCMHs does not exceed the existing  limits, we will derive the upper limits on the abundance of PBHs. 

This paper is organized as follows. In Sec. II we investigate the basic properties of UCMHs, and then derive the upper limits on the abundance of PBHs. The conclusions are given in Sec. III. Throughout the paper we will use the cosmological parameters from Planck-2018 results~\cite{planck-2018}.

%%%%%%%%%%%%%%%%%%%%%%%%%%%%%%%%%%%%%%%%%%%%%%%

\section{The basic properties of UCMHs and upper limits on the abundance of PBHs}

The formation mechanism of UCMHs has been studied in previous works, see, e.g., Refs.~\cite{0908.0735,Eroshenko:2016yve,Adamek:2019gns,Boucenna:2017ghj}. There are two possible mechanisms: (i) direct collapse of large density perturbation ($10^{-4}\lesssim \delta \rho/\rho \lesssim 0.3$) in the early universe, which is not large enough to form PBHs; (ii) accretion of DM particles onto PBHs. We will not discuss the formation mechanism of UCMHs and adopt the second scenario for our purpose. 

In the radiation dominated early universe, a large density perturbation, $\delta \rho/\rho \gtrsim 0.3$, can directly collapse to form PBHs~\cite{carr,pbhs_review}. Then DM particles can be accreted onto PBHs resulting in the formation of UCMHs. The mass of a UCMH changes slowly during the radiation dominated period and increases significantly after the redshift of matter-radiation equality $z_{\rm eq}$~\cite{0908.0735}. The specific form of the changes of mass with redshift is~\cite{0908.0735,ucmh_scott} 

\beqa
M_{\rm UCMH} = \left(\frac{1+z_{\rm eq}}{1+z}\right)M_{\rm PBH}
\label{eq:m_ucmh}
\eeqa 

The density profile of DM in a UCMH is in the form of~\cite{ucmh_scott,dongzhang} 

\beqa
\rho_{\rm DM}(r) = \frac{3M_{\rm UCMH}(z)}{16\pi R_{\rm UCMH}(z)^{3/4}r^{9/4}},
\label{eq:rho_ucmh}
\eeqa
where $R_{\rm UCMH}(z)$ is the radius of UCMH~\cite{0908.0735}, 

\beqa
R_{\rm UCMH}(z) = 0.019~{\rm pc}\left(\frac{1000}{1+z}\right)\left(\frac{M_{\rm UCMH}(z)}{M_{\odot}}\right)^{1/3}
\label{eq:r_ucmh}
\eeqa 

For the density profile shown in Eq.~(\ref{eq:rho_ucmh}), $\rho_{\rm DM}(r)\to \infty$ for $r\to 0$. Taking into account the annihilation of DM particles, there is a maximum core density, $\rho_{\rm  max}$, at the center of a UCMH~\cite{ucmh_scott,epjplus-1}, 

\beqa
\rho_{\rm max} = \frac{m_{\chi}}{\left<\sigma v\right>(t-t_{i})},
\label{eq:rho_max}
\eeqa
where $t_i$ is the formation time of UCMH and here we set $t_{i}=t_{\rm eq}$. $m_{\chi}$ and $\left<\sigma v\right>$ are the mass and thermally averaged annihilation cross section of DM particles, respectively. The corresponding core radius, $r_{\rm core}$, is defined as $\rho_{\rm max}=\rho_{\rm DM}(r_{\rm core})$. In short, the density profile of the DM particle within a UCMH used here is as follows:

\beqa
\rho(r)=\left\{
\begin{array}{rcl}
\rho_{\rm max} &&	{r\le r_{\rm core}}\\
\rho_{\rm DM}(r) && {r\ge r_{\rm core}}\\
\end{array} \right. 
\eeqa

The annihilation rate of the DM particle within a UCMH can be written as 

\beqa
\Gamma_{\rm anni}&&=\int_{0}^{R_{\rm UCMH}} 2\pi n^{2}_{\rm DM}(r)\left<\sigma v\right>r^{2}dr\nonumber \\
&&=2\pi\frac{\left<\sigma v\right>}{m_{\chi}^2}\int_{0}^{R_{\rm UCMH}}\rho_{\rm DM}^{2}(r)r^{2}dr \nonumber \\
&&=2\pi\frac{\left<\sigma v\right>}{m_{\chi}^2}\left[\int_{0}^{r_{\rm core}}\rho_{\rm max}^{2} + 
\int_{r_{\rm core}}^{R_{\rm UCMH}}\rho_{\rm DM}^{2}\right]r^{2}dr
\eeqa

In the redshift range $1100\lesssim z\lesssim 5\times10^{4}$, the heated electrons produced by Compton scattering or pair production on ions transfer their energy by inverse Compton scattering to the CMB photons, producing a distorted spectrum with phase-space distribution~\cite{McDonald:2000bk,Chluba:2011hw,waynehu}, 

\beqa
f(x,y) \simeq f(x,0)+ y\frac{xe^{x}}{(e^{x}-1)^{2}}\left[\frac{x}{{\rm tanh}(x/2)}-4\right],
\label{eq:f}
\eeqa
where $f(x,0)=1/(e^{x}-1)$ is the Planck distribution with $x=E/T$. $y$ is the parameter used to measure the strength of CMB $y$-type distortion. The distribution function $f(x,y)$ is governed by the Kompaneets equation~\cite{Kompaneets}, and after some algebraic calculations one can find the relation between the parameter $y$ and the energy injected into the Universe~\cite{McDonald:2000bk,Chluba:2011hw}, 

\beqa
y = \frac{1}{4}\frac{\delta \rho_{\gamma}}{\rho_{\gamma}},
\label{eq:y}
\eeqa
where $\delta \rho_{\gamma}$ is the injected energy and $\rho_{\gamma}$ is the energy density of the CMB photons. 

The energy injection rate of DM annihilation per unit volume can be written as

\beqa
E_{\rm inj}=n_{\rm PBH}m_{\chi}\Gamma_{\rm anni}  = \Omega_{\rm PBH}\rho_{c,0}\frac{m_{\chi}}{M_{\rm PBH}}\Gamma_{\rm anni},
\label{eq:e_inj}
\eeqa
where $n_{\rm PBH} = \rho_{\rm PBH}/M_{\rm PBH}$ is the number density of PBH, 
$\rho_{c,0}$ is the critical density of the Universe at $z=0$ and $\Omega_{\rm PBH}=\rho_{\rm PBH}/\rho_{c,0}$. In this paper, we have adopted a monochromatic PBH mass function for our calculations.

Using the above equations, the parameter $y$ caused by DM annihilation within UCMHs can be rewritten as follows:

\beqa
y_{\rm DM}&&= \frac{1}{4}\frac{\delta \rho_{\gamma}}{\rho_{\gamma}}=
\frac{1}{4}\int^{t_2}_{t_{1}}\frac{{\dot{\rho}_{\rm anni}}}{\rho_{\gamma}}dt 
= \frac{1}{4}\int^{t_2}_{t_{1}}\frac{E_{\rm inj}(z)}{\rho_{\gamma 0}(1+z)^{4}}dt \nonumber \\
&&=\frac{1}{4}\frac{1}{\rho_{\gamma 0}}\int^{z_{\rm eq}}_{z_{\rm rec}}\frac{E_{\rm inj}(z)}{H(z)(1+z)^{5}}dz,
\label{eq:y_int}
\eeqa
where $H(z)=H_{0}\sqrt{\Omega_{\rm m}(1+z)^{3}+\Omega_{\Lambda}+{\Omega_{\gamma}}(1+z)^{4}}$, 
$\rho_{\gamma 0}$ is the energy density of CMB at $z=0$. Here we have used the relation $dt = 1/H(z)(1+z)~dz$

\begin{figure}
\centering
\includegraphics[width=0.5\textwidth]{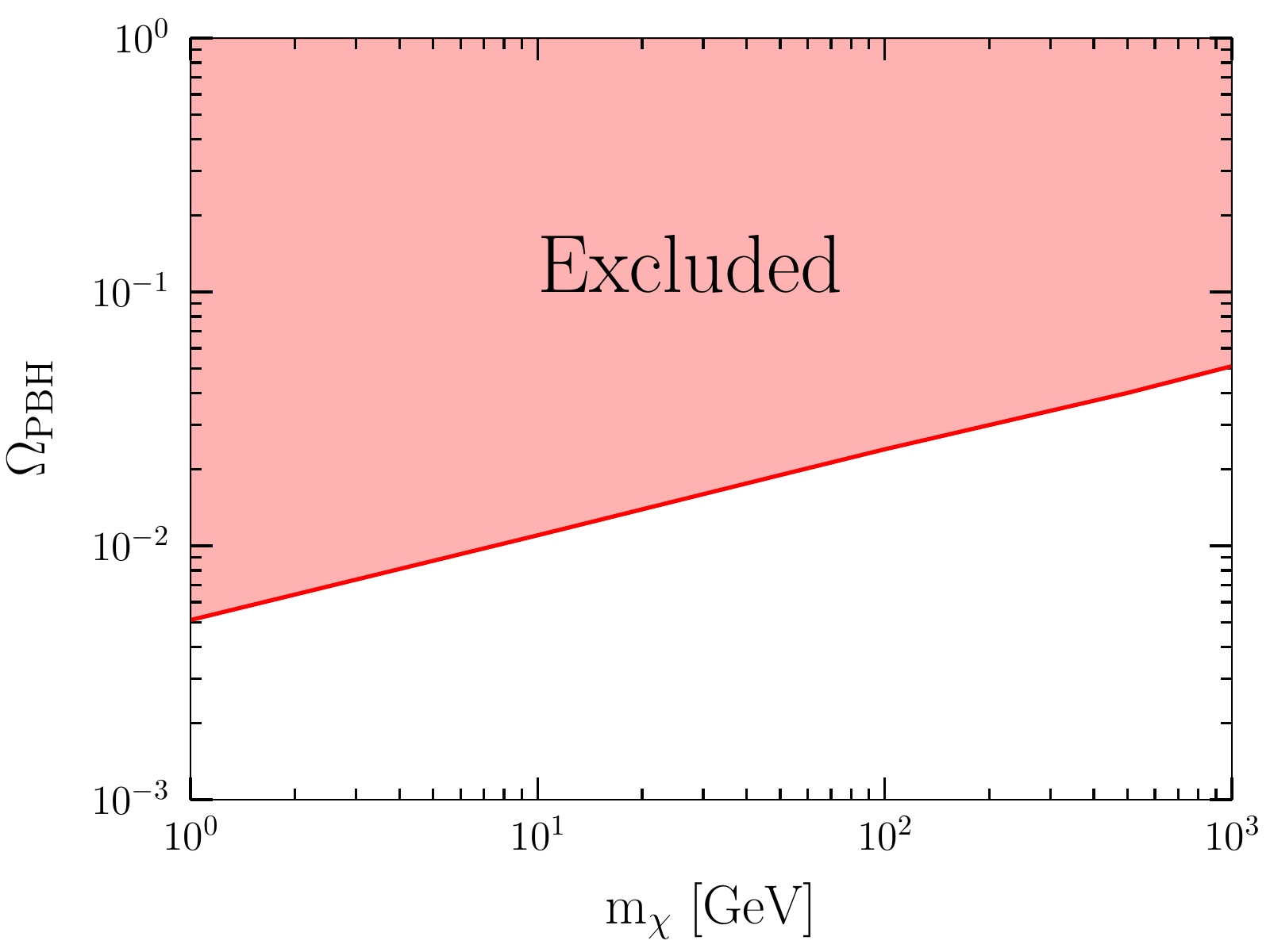}
\caption{Upper limits on the abundance of PBHs, $\Omega_{\rm PBH}=\rho_{\rm PBH}/\rho_{c,0}$, by requiring $y_{\rm DM}<1.5\times10^{-5}$. Here we have set the thermally averaged annihilation cross section 
$\left<\sigma v\right>=3\times 10^{-26}~\rm cm^{3}~s^{-1}$.}
\label{fig:fraction}
\end{figure}

Using Eq.~(\ref{eq:y_int}) and by requiring $y_{\rm DM} <1.5\times10^{-5}$, we derive the upper limits on the abundance of PBHs and the results are shown in Fig.~\ref{fig:fraction}. From this plot, it can be found that for the DM mass range $1\le m_{\chi}\le 10^{3}~\rm GeV$ with $\left<\sigma v\right>=3\times 10^{-26}~\rm cm^{3}~s^{-1}$, the upper limits on the abundance of PBHs are $5\times 10^{-3}\le \Omega_{\rm PBH}\le 5\times 10^{-2}$, and a fit to the boundary line is as follows: 

\beqa
\Omega_{\rm PBH} \le 5\times 10^{-3}~\left(\frac{m_{\chi}}{\rm GeV}\right)^{\frac{1}{3}}. 
\eeqa

Note that the limits are independent on the PBH mass, and similar features can be also found in previous works~\cite{Zhang:2021mth,ucmh_scott,Yang:2011jb,Yang_2011,Kadota:2022cij}. In fact, since the annihilation rate of the DM particle within UCMH is proportional to the PBH mass $\Gamma_{\rm anni}\propto M_{\rm PBH}$, the energy injection rate per unit volume is $E_{\rm inj}\propto n_{\rm PBH}\Gamma_{\rm anni}\propto M_{\rm PBH}^{-1}M_{\rm PBH}$. As shown in Fig.~\ref{fig:fraction}, the limits are stronger for lower DM mass due to the larger number density.

%%%%%%%%%%%%%%%%%%%%%%%%%%% Figure 1 %%%%%%%%%%%%%%%%%%%%%%%%%%%%%%%

%%%%%%%%%%%%%%%%%%%%%%%%%% Figure 1 %%%%%%%%%%%%%%%%%%%%%%%%%%%%%%%%%

\begin{figure}
\centering
\includegraphics[width=0.5\textwidth]{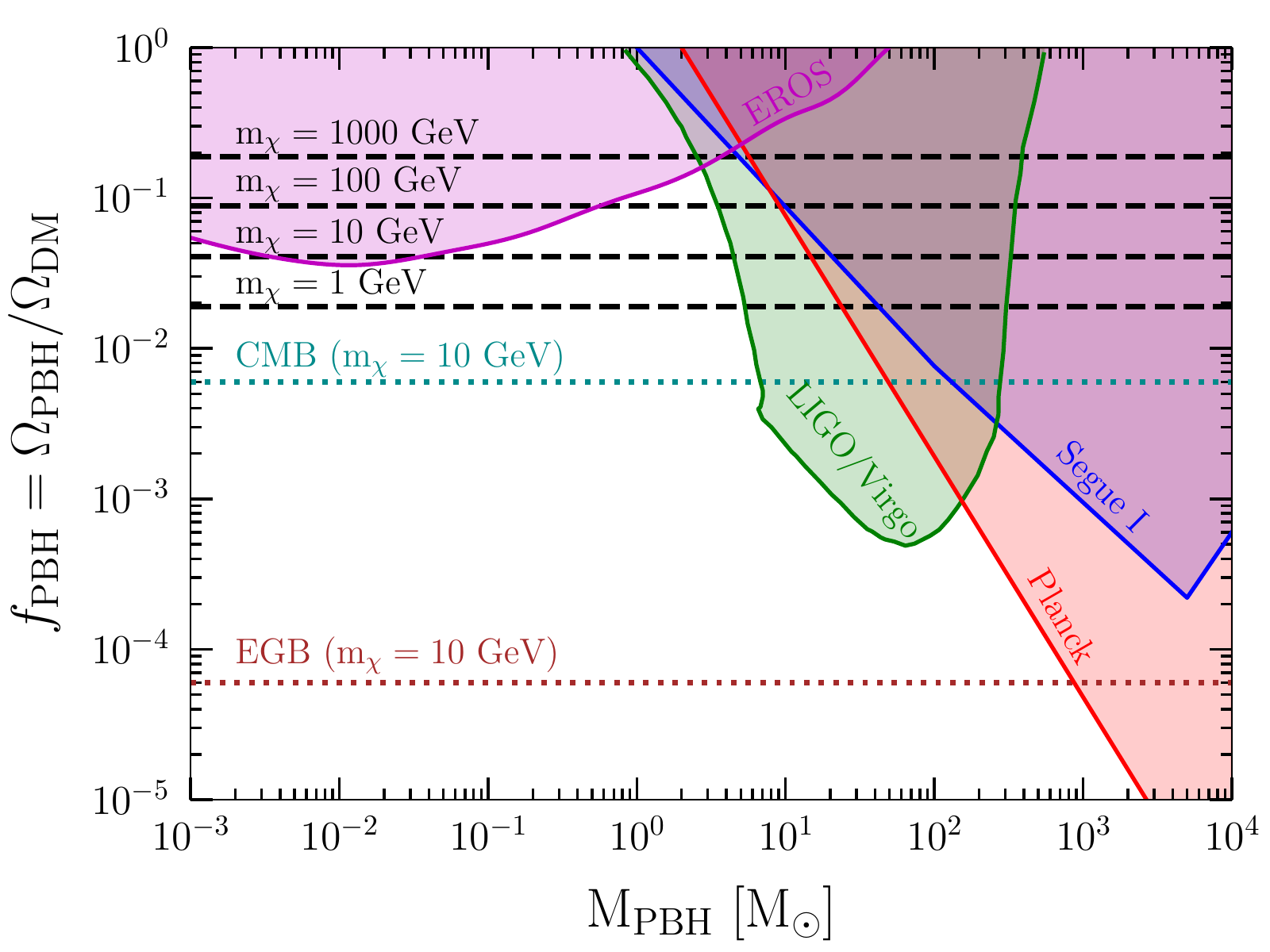}
\caption{Comparison with a few other constraints on the fraction of DM in PBHs, $f_{\rm PBH}=\Omega_{\rm PBH}/\Omega_{\rm DM}$. The horizontal dashed lines show the upper limits obtained by us for the DM mass $m_{\chi}=1,10,100$ and 1000 GeV. Other constraints are from: 1) the merger rate of PBHs in light of the sensitivity of LIGO/Virgo (labeled 'LIGO/Virgo')~\cite{Vaskonen:2019jpv} (updated constraints from the O3 observational run can be found in Ref.~\cite{Hutsi:2020sol}); 2) the influence of PBHs on the dynamical evolution of stars in the dwarf galaxy Segue I (labeled ‘Segue I’)~\cite{segue}; 3) investigating the influence of accreting PBHs on the CMB with Planck data (labeled ‘Planck’)~\cite{Poulin:2017bwe}; 4) the gravitational lensing effects based on EROS (labeled ‘EROS')~\cite{EROS}; 5) the contribution of DM annihilation within UCMHs to the EGB (labeled ‘EGB')~\cite{Zhang:2021mth};and 6) the influence of DM annihilation within UCMHs on the CMB (labeled ‘CMB')~\cite{Yang:2011jb}.} 
\label{fig:fraction_compare}
\end{figure}

For comparison, the upper limits on the fraction of DM in PBHs, $f_{\rm PBH}=\Omega_{\rm PBH}/\Omega_{\rm DM}$, from several other  observations are shown in Fig.~\ref{fig:fraction_compare} (here $\Omega_{\rm DM}=\Omega_{\rm WIMPs} + \Omega_{\rm PBHs}$). From this plot, it can be seen that our limits provide a useful complement for the mass range $10^{-3}\lesssim M_{\rm PBH}\lesssim 1~M_{\odot}$, where the limits are obtained mainly by investigating the gravitational lensing effects based on the results of the European Southern Observatory (EROS)~\cite{EROS}. 

%%%%%%%%%%%%%%%%%%%%%%%%%%%%% Figure 2 %%%%%%%%%%%%%%%%%%%%%%%%%%%
 
%%%%%%%%%%%%%%%%%%%%%%%%%%%%% Figure 2 %%%%%%%%%%%%%%%%%%%%%%%%%%%

In the mixed dark matter scenarios investigated here the observations of $\gamma$-ray and CMB can be also used to constrain the  abundance of PBHs. For example, the high energy photons produced by the DM annihilation within UCMHs can contribute to the extragalactic $\gamma$-ray background (EGB)~\cite{Zhang:2021mth,ucmh_scott,Yang:2011jb,Yang_2011,epjplus-2,Boucenna:2017ghj,Carr:2020erq,Carr:2020mqm,Nakama:2017qac,Gines:2022qzy,Belotsky:2005dk}. Using the observations of EGB by, e.g., the Fermi telescope, the upper limits on the abundance of PBHs can be obtained~\cite{Zhang:2021mth,Yang_2011,epjplus-2}. The authors of~\cite{Zhang:2021mth} used the updated Fermi-LAT EGB measurement to derive the upper limits on the abundance of PBHs. They found that for the $b\bar b$ annihilation channel the upper limit is $f_{\rm UCMH}\sim 6\times 10^{-5}$ for $m_{\chi}=10~\rm GeV$ with $\left<\sigma v\right>=3\times 10^{-26}~\rm cm^{3}~s^{-1}$ (shown in Fig.~\ref{fig:fraction_compare}, horizontal dotted line labeled `EGB'). After recombination, the energy released from DM annihilation within UCMHs can be injected into the intergalactic medium (IGM), leading to the changes of the IGM thermal history~\cite{Yang:2011jb,dongzhang,epjplus-1}. These changes can be constrained by, e.g., the CMB observations. In Ref.~\cite{Yang:2011jb}, the authors have used the CMB data to get the upper limits on the abundance of PBHs. They found that the upper limit is $f_{\rm UCMH}\sim 6\times 10^{-3}$ for $m_{\chi}=10~\rm GeV$ with $\left<\sigma v\right>=3\times 10^{-26}~\rm cm^{3}~s^{-1}$ (shown in in Fig.~\ref{fig:fraction_compare}, horizontal dotted line labeled `CMB').  

\section{conclusions}

We have investigated the abundance of PBHs in the mixed dark matter scenarios consisting of PBHs and WIMPs. In this scenario, UCMHs can be formed around PBHs with a density profile $\rho(r)\sim r^{-9/4}$. The annihilation rate of particle DM (WIMPs) within UCMHs is enhanced and has influence on the evolution of the Universe. Previous works have focused these influences on the epoch after recombination. Since UCMHs can be formed in the early universe, they can effect CMB $y$-type spectral distortion in the redshift range $z_{\rm rec} \le z\le z_{\rm eq}$. We have investigated the influence of DM annihilation within UCMHs on the CMB $y$-type distortion. By requiring that the parameter $y$ caused by DM annihilation within UCMHs does not exceed the current limits $|y| < 1.5\times 10^{-5}$, we derived the upper limits on the abundance of PBHs, $5\times 10^{-3}\le \Omega_{\rm PBH}\le 5\times 10^{-2}$ for the WIMPs mass range $1\le m_{\chi}\le 1000~\rm GeV$. Compared with other existing limits, although our limits are not the strongest, they are new results obtained in a different way.

\section{Acknowledgements}
This work is supported by the Shandong Provincial Natural Science Foundation (Grant No.ZR2021MA021). 
Y. Yang is supported in part by the Youth Innovations and Talents Project of Shandong Provincial Colleges and Universities (Grant No. 201909118).
\

\bibliographystyle{apsrev4-1}
\bibliography{refs}

\end{document}